\documentclass{elsart}

\usepackage{epsf,epsfig}

\usepackage{amssymb}

\newcommand{\geteps}[2]{\epsfig{figure=PS/#2,width=#1}}
\newcommand{\beq}{\begin{equation}}
\newcommand{\eeq}{\end{equation}}
\newcommand{\beqar}{\begin{eqnarray}}
\newcommand{\eeqar}{\end{eqnarray}}
\newcommand{\del}{\partial}                     
\newcommand{\K}{K}
\newcommand{\gsim}{\mbox{\raisebox{-0.6ex}{$\stackrel{>}{\sim}$}}\:}
\newcommand{\lsim}{\mbox{\raisebox{-0.6ex}{$\stackrel{<}{\sim}$}}\:}
\def\GeV{{\rm GeV}}

\begin{document}

\begin{frontmatter}



\title{
{\large Sideward Peak of Intermediate Mass Fragments \\
in High Energy Proton Induced Reactions}
}


\author[Hokudai,JAERI-Adv,Genome]{Yuichi Hirata},
\author[Hokudai]{Akira Ohnishi},
\author[BNL]{Yasushi Nara},
\author[JAERI-Adv,Hitechs]{Toshihiko Kido},
\author[JAERI-Adv]{Toshiki Maruyama},
\author[Hokudai,Hokudai-Eng]{Naohiko Otuka},
\author[RIST]{Koji Niita},
\author[JAERI-Neu]{Hiroshi Takada}, 
\author[JAERI-Adv]{Satoshi Chiba}

\address[Hokudai]{
	Division of Physics,
	Graduate School of Science, \\
	Hokkaido University,
	Sapporo 060-0810, Japan}
\address[BNL]{
	RIKEN BNL Research Center, \\
	Brookhaven National Laboratory,
	Upton, NY 11973, USA}
\address[RIST]{
	Research Organization for Information Science $\&$ Technology, \\
	Ibaraki 319-1195, Japan}
\address[JAERI-Adv]{
	Advanced Science Research Center, \\
	Japan Atomic Energy Research Institute,
	Ibaraki 319-1195, Japan}
\address[JAERI-Neu]{
	Center for Neutron Science, \\
	Japan Atomic Energy Research Institute,
	Ibaraki 319-1195, Japan}
\address[Genome]{
	Human Genome Center,
	Institute of Medical Science,\\
	University of Tokyo,
	Tokyo 108-8639, Japan}
\address[Hokudai-Eng]{
        Meme Media Laboratory, 
        Graduate School of Engineering,\\
        Hokkaido University,
        Sapporo 060-8628, Japan}
\address[Hitechs]{
        Hitechs Inc., Kurobe 938-0044, Japan
        }              

\begin{abstract}
Intermediate mass fragment (IMF) formation
in the 12 GeV proton induced reaction on Au target is analyzed
by using a combined framework of a transport model (JAM/MF)
and a newly developed non-equilibrium percolation (NEP) model.
In this model, we can well reproduce 
the mass distribution of fragments.
%
In addition, 
the sideward peaked angular distribution would emerge 
under the condition that the fragment formation time 
is very short, around 20 fm$/c$.
Within this short time period,
the un-heated part of the residual nucleus is kept to have 
doughnut shape, then the Coulomb repulsion from this shape 
strengthens the sideward peak of IMF.
%
\end{abstract}

\end{frontmatter}

\section{Introduction}
\label{sec:Introduction}

%
%
Multifragment formation from excited nuclei has attracted attention
in this decade.
It is closely related to the properties of nuclear matter, 
especially to its liquid-gas phase transition.
For example, it has been proposed that 
multifragmentation would be caused by the volume instability 
of nuclear matter in the spinodal region~\cite{Guide,Lynch:1987ib},
in which nuclear liquid and gas phase rapidly co-exist.
The above idea on the fragment formation has received renewal interest,
since recent high-quality heavy-ion experiments provide us of strong evidences 
of the first-order nature of 
the nuclear liquid-gas phase transition~\cite{CaloricCurve,NegativeHC,FOPI}.
It is a big challenge for theorists to elucidate the relation between 
the fragment formation and the equilibrium properties of nuclear matter,
including its phase transitions.
In heavy-ion collisions, however, non-trivial roles of strong collective flow
make it difficult to relate fragment formations and the nuclear matter
properties directly.
Thus it is also necessary to investigate
fragment formation phenomena in light-ion induced reactions,
in which the flow effects on the fragmentation are expected to be smaller.

%
%
It has been known that the fragment mass distribution 
in high-energy proton induced reactions follows the power law,
which signals the statistical formation of fragments
in a critical but equilibrium condition.
Energy spectra of IMFs also support the statistical fragment formation picture.
In a recent KEK experiment of 12 GeV proton induced reaction on Au target, 
the energy spectra of IMFs such as O, F, Ne, and Na are measured,
and it is found to be well fitted with a thermal model with Coulomb barrier.

%
%
Contrary to the mass and energy distribution of fragments,
the angular distribution of IMFs contradicts
to a picture of the statistical fragmentation from an equilibrated system.
It is also long known that the angular distribution of
Intermediate Mass Fragments (IMFs) becomes sideward peaked
at proton incident energies above $E_p =$ 10 GeV~\cite{KEK-pAu}.
In the above KEK experiments, 
the angular distribution of IMFs such as O, F, Ne, and Na
are shown to have a sideward peak around 70 degrees in the laboratory frame.
The same experimental conclusion is also obtained in a BNL-AGS experiment.
In the AGS p+Au experiment~\cite{AGS}, the laboratory angular distribution 
of C isotopes becomes to have sideward peak 
as the incident energy of proton go beyond 10 GeV.
%
%
This sideward peaked feature has been considered to be a mystery;
it contradicts to a naive picture of statistical fragment formation 
from a spherical equilibrated residual nuclei.

In a statistical picture of fragment formation, 
the angular distribution of fragments is generally forward peaked.
At low energies, the incident proton is absorbed
into the target, and forms a thermalized compound nucleus
having the incident proton momentum.
At high energies, 
the incident proton collides with nucleons inside target nucleus
(cascade process) in the first stage of reactions.
These successive two-body scatterings give
momentum transfers to the nucleons in target,
and some of them are emitted from the residual nucleus. 
The momentum transfer to the residual nucleus
is in the forward direction in average.
Thus in both cases, the residual excited nucleus
should have a forward momentum in average.
If this residual compound nucleus decays statistically, 
produced fragments should have forward peaked angular distribution.

%
%
It is expected that these fragment angular distribution anomaly 
would be due to a new fragment formation mechanism, 
which becomes evident at higher incident energies.
This new mechanism would be related to a new property
of excited nuclei produced in GeV proton induced reactions.
For the understanding of this sideward emission of IMFs,
several characteristic mechanisms 
have been proposed so far;
the shock wave propagation~\cite{Shock-Wave}, 
the strange shaped nucleus formation~\cite{BUU-pA,AlphaAu-Maru}
which is also predicted to be formed
in heavy ion collisions~\cite{Bubble1,Bubble2},
and recoil effects of particle scatterings~\cite{AGS}.
However,
there is neither decisive conclusion nor satisfactory explanation 
for the sideward peak of IMF angular distribution,
since it is not an easy task to analyze this phenomenon theoretically.

%
%
In the first stage of the high-energy proton induced reaction,
the incident proton interacts with target nucleons and produces several pions.
In the incident energy range around 10 GeV,
the reaction mechanism of hadron-hadron collision 
evolves from single resonance baryon production
to mutual excitation of incident baryons and/or string formation.
For a reliable description of high-energy $pA$ reactions, 
it is necessary to invoke these incident-energy dependent 
elementary cross sections parametrized and verified in $pp$ and $pA$ reactions.
In Ref.~\cite{JAM}, a hadron-string cascade model (JAM) is developed,
and it is successfully applied to $hh$ and $pA$ reactions
as well as $AA$ reactions.
%
%
In the next stage of the reaction,
scattered nucleons evolve in the mean field
and then form an excited residual nuclei.
Since the nucleon kinetic energy ranges from a few MeV to around 10 GeV
and the fragment mass ranges from two (deuteron) to around 200,
it is necessary to invoke appropriate mean field 
which takes account of the energy dependence as well as the density
dependence and is verified in $pA$ reactions.
In Ref.~\cite{QMD-MEANB}, 
a parameterization of nuclear mean field is proposed, 
and this mean field is verified in $pA$ reactions in a wide range
within a framework of the Quantum Molecular Dynamics (QMD).
In addition, it is already shown to well describe
the incident energy dependence of the nucleon potential depth
and the binding energy behavior from light to the heaviest nuclei.
%
%
In the third stage of the reaction,
the residual nucleus breaks up into fragments having various masses
including IMFs, and the fragmentation mechanism is not understood well
in the case we cannot assume that the system decays
statistically from an equilibrated system.
In heavy-ion reactions,
there are some dynamical transport models which do not assume 
equilibrium~\cite{Non-eq,OR97b,OH96c} and well reproduce the IMF multiplicities.
In this case, the system expands to sub-saturation density
due to the rapid collective expansion 
from the initial hot and compressed nuclei.
At sub-saturation densities, local density fluctuation is energetically
favored, then fragments are easily formed in the dynamical stage.
However, in the case of $pA$ reactions, the flow effect is not large and
the expansion is not enough to put the system into sub-saturation densities.
At around the saturation density, 
density fluctuation is not energetically favored,
but smoothed out during the evolution in a usual mean field type dynamics
even if it is generated in the first stage of the reaction.
Thus at present, there is no satisfactory microscopic model to describe 
the non-equilibrium fragmentation process 
in high energy proton induced reactions.

%
%
In order to develop a model of non-equilibrium fragmentation 
in proton-induced reactions, we start from an equilibrium fragmentation
model, percolation.
It has been shown that the percolation model gives a good description
of mass and momentum distribution of fragments in high-energy proton 
induced reactions, if we ignore the anisotropy of fragment emission
in the residual nuclear rest frame.
Since the initial nucleons are assumed to be distributed uniformly in a sphere
in the equilibrium percolation model, the fragment angular distribution
becomes automatically isotropic in the residual nuclear rest frame.
Therefore, it is expected that the fragment angular distribution would be
anisotropic if the initial nucleon distribution contains dynamical information
of the earlier stage. 
We here formulate a new Non-Equilibrium Percolation (NEP) model
which takes account of initial density and momentum fluctuation 
generated in the earlier dynamics.
In NEP, the nucleon phase space variables are taken from the results
of the transport model calculation as the initial condition of percolation,
instead of putting nucleons on sites.
The bond breaking or connection probability is calculated
by using the distance and momentum difference between the two nucleons
under consideration, instead of giving a common bond breaking probability
for the bonds connecting nearest neighbor sites.
By doing these modifications, 
we can take into account the position and momentum fluctuations 
of nucleons inside fragment source
and study the effects of non-equilibrium features of residual nucleus
to the IMF formation processes.

%
%
In this paper, we study the sideward enhanced IMF formation mechanism 
by using a combined framework of a transport model
and a newly developed Non-Equilibrium Percolation model.
In the transport part,
we use the cross sections and particle production algorithm developed
in JAM~\cite{JAM},
and the mean field verified in QMD~\cite{QMD-MEANB}.
Hereafter we call this transport model, JAM/MF (JAM with mean field).
The nucleon phase space information given in the transport model
is used as the initial condition of NEP.
We find that
we can well reproduce the mass and energy distributions of IMFs
in a unified way by using the above combined framework.
In addition, if we assume that fragments are produced within a short time
scale, we can also describe the sideward enhancement of IMF emission.
%

%
%
We find the following reaction mechanism to enhance sideward emission of IMF:
In the short time scale (around 20 fm$/c$), 
the dynamical effect in the first hadronic cascade stage still remains.
Nucleons along the path of the leading incident proton are heated-up
and have large momenta in average.
Since IMFs are formed mainly in the cold region, 
the probability of IMF formation is small around the leading proton path.
Then the IMF formation points are distributed non-spherically,
predominantly distributed in a doughnut shape.
After forming fragments in this doughnut shape, 
Coulomb repulsion between fragments pushes more strongly in the sideward
direction in the rest frame of the residual nuclei.

%
%
The outline of this paper is as follows.
In Sec.~\ref{sec:Transport}, 
we explain the transport model JAM/MF used in this work.
In Sec.~\ref{sec:NEP}, 
a newly developed Non-Equilibrium Percolation (NEP) model is explained,
and we discuss the validity of this model
to show the good reproduction of fragment mass distribution.
Next in Sec.~\ref{sec:Results}, 
we discuss the mechanism of sideward enhancement of IMF emission
through the analysis of the production position distribution of IMF,
the IMF energy distribution and the IMF angular distribution.
Finally in Sec.~\ref{sec:Summary}, we summarize this paper.

\section{Transport Model for High Energy Proton Induced Reactions}
\label{sec:Transport}

In high-energy ($E_p \gsim 10 \GeV$) proton induced reactions, 
there are mainly three stages
--- hadronic cascade, mean field evolution, and fragmentation.
During these stages, nucleon energy decreases
by about four order of magnitude from around 10 GeV to a few MeV.
A large part of the proton incident energy ($\sim$ 10 GeV) is carried away 
by produced pions and emitted high momentum nucleons, 
and only a part of this energy remains as the excitation energy of the residual
nucleus (several hundred MeV to a few GeV). 
This residual nuclear excitation energy is mainly exhausted 
to breakup the residual nucleus into fragments and nucleons,
then the energy of finally produced fragments are around a few MeV per nucleon.

For a reliable microscopic description of dynamical evolution of the system
throughout these stages,
it is necessary to apply 
	reliable hadronic cross sections for hadronic cascade processes,
and 
	well verified mean field. 
%
%
In this work,
we have developed a transport model, JAM/MF,
by extending the hadron-string cascade model JAM~\cite{JAM}
to include the mean field developed in Ref.~\cite{QMD-MEANB}.
It can be also considered as a version of QMD
with hadronic cross sections developed in JAM.
%
%
%
%
%
%
The results of JAM/MF are used as the initial configuration
of the Non-Equilibrium Percolation described in the next section.

\subsection{Hadronic cascade processes in JAM}
\label{subsec:JAM}

%
%
Here we describe the outline of the hadron-string cascade model, JAM.
The detail of JAM is described in Ref.~\cite{JAM}.
The main components of JAM are as follows.
\begin{enumerate}
\item
    Nuclear collision is assumed to be described by the
    sum of independent binary $hh$ collisions.
    Each $hh$ collision is realized by the closest distance approach.
    In the original version of JAM, no mean field effect is included, 
    therefore the trajectory of each hadron is straight in between two-body
    collisions, decays or absorptions.
\item
    The initial position of each nucleon is sampled by the
    parameterized distribution of nuclear density.
    Fermi motion of nucleons are assigned
    according to the local Fermi momentum.
\item
    All established hadronic states, 
    including resonances with masses up to around 2 GeV,
    are explicitly included
    with explicit isospin states as well as their anti-particles.
    All of them can propagate in space-time.
\item
    The inelastic $hh$ collisions produce resonances at low energies
    while at high energies
    ($\sqrt{s} \gsim 4\GeV$ in $BB$ collisions 
     $\sqrt{s} \gsim 3\GeV$ in $MB$ collisions
    and $\sqrt{s} \gsim 2\GeV$ in $MM$ collisions)
    color strings are formed and they decay into hadrons according to the
    Lund string model~\cite{PYTHIA}.
    Formation time is assigned to hadrons from string fragmentation.
    Formation point and time are determined by assuming yo-yo formation point.
    This choice gives the formation time of
    roughly 1 fm$/c$ with string tension $\kappa=1 $GeV/fm.
\item
    Hadrons which have original constituent quarks can scatter
    with other hadrons assuming the additive quark cross section
    within a formation time.
    The importance of this quark(diquark)-hadron interaction
    for the description of baryon stopping
    at CERN/SPS energies was reported by Frankfurt group~\cite{rqmd1,urqmd}.
\item
    At very high energies ($\sqrt{s} \gsim 10\GeV$),
    multiple minijet production is also included in the same way
    as the HIJING model~\cite{HIJING} in which jet cross section and the number
    of jet are calculated using an eikonal formalism for
    perturbative QCD (pQCD).
    Hard parton-parton scatterings
    with initial and final state radiation are simulated
    using PYTHIA~\cite{PYTHIA} program.
\item
    Pauli-blocking for the final nucleons in two-body collisions
    are also considered.
\item
    We do not include any medium effect
    such as string fusion to rope~\cite{venus,rqmd1},
    medium modified cross sections and in-medium mass shift.
    All results which will be presented in this paper
    are those obtained from the free cross sections and free masses as inputs.
\end{enumerate}

%
%
Hadron (proton, $\pi^+, \pi^-, \K^+, \K^-$) transverse mass 
$m_t$ in proton induced reactions (p(14.6GeV/c) + Be, Al, Cu, Au), 
	light-heavy ion induced reactions (Si(14.6A GeV/c)+ Al, Cu, Au), 
and	heavy-ion reactions (Au(11.6A GeV/c) + Au)
has been analyzed in JAM,
and it is found that JAM results show good agreement with the experimental data
without any change of model parameters~\cite{JAM}.
This means that momentum distributions of isolated single proton are correctly 
calculated by collision cross sections of JAM.

On the other hand, 
it is required to take account of later mean field evolution and percolation
processes for the description of IMF production.

\subsection{Propagation in Quantum Molecular Dynamics}
\label{subsec:QMD}

In the framework of Quantum Molecular Dynamics (QMD),
the classical trajectory of hadrons is determined 
by the stochastic hadronic collisions, described in the previous subsection,
and the Newtonian equation of motion.
The single particle wave function of each nucleon is represented
by a Gaussian wave packet, having the phase space centroid parameter
of ($\bold{R}_i, \bold{P}_i$) for the $i$-th nucleon.
The total wave function is assumed to be a product wave function
of nucleon Gaussian wave packets.
Then we can derive the Newtonian equations of motion 
on the basis of the time dependent variational principle~\cite{QMD-ORG}.
The equations of motion become the canonical equations
of the Gaussian centroid parameters as phase space variables, 
\beq
\dot{\bold{R}}_k =  \frac{\del H}{\del \bold{P}_k} \ ,\quad
\dot{\bold{P}}_k = -\frac{\del H}{\del \bold{R}_k} \ .
\eeq
The Hamiltonian $H$ consists of the kinetic energy
and the effective interaction energy.
\beqar
\label{eq:thamil}
H &=& T + V 
\ ,\\
T &=& \sum_i E_i = \sum_i \sqrt{m^2_i + \bold{P}^2_i}
\ ,\\
\label{eq:effect}
V &=&
    V_{\rm{Skyrme}}
  + V_{\rm{Sym}}
  + V_{\rm{Mom}}
  + V_{\rm{Coul}}
  + V_{\rm{Pauli}}
\ .
\eeqar
In this interaction energy,
the following terms are included:
	Skyrme type density dependent interaction ($V_{\rm{Skyrme}}$),
	symmetry energy ($V_{\rm{Sym}}$), 
	momentum dependent interaction taking account of
		the energy dependence of the nucleon potential 
		($V_{\rm{Mom}}$), 
	Coulomb potential between protons ($V_{\rm{Coulomb}}$),
	and the Pauli potential ($V_{\rm{Pauli}}$).
The last term is added to take into account the Pauli principle approximately.
The form of each term is shown in the Appendix.

In this study we use parameters shown in Table~\ref{tab:QMDpara}.
These parameters are determined to reproduce
various kinds of systematic observables such as 
	(1)the density dependence of the total kinetic energy of Fermi gas, 
	(2)density dependence of the total energy of symmetric nuclear matter
		above normal nuclear density,
	and 
        (3)energy dependence of the real part of proton-nucleus optical 
                potential,
	        binding energy and radius of heavy nuclei ($A > 3$). 
All these observables are calculated by metropolis sampling method 
under the condition that the temperature of nucleus is 2.5 MeV.
Including the Pauli potential improves the reproductivity of 
the quasi-elastic part of the neutron energy spectra in
p(256MeV,800MeV)+\nuc{208}{Pb}~\cite{QMD-MEANB}.

In the simulation of 12 GeV proton induced reaction,
the ground state of target nucleus (Au) is calculated by frictional 
cooling and heating method to fit the binding energy of Au. 

In the total Hamiltonian $H$ we have already introduced the relativistic
form of kinetic energy expression and adopted the relativistic kinematics
in the cascade processes in JAM.
In order to make a full relativistic description,
it is also necessary to formulate the interaction terms in a covariant way.
A Lorentz-covariant extension of the QMD,
debbed relativistic quantum molecular dynamics (RQMD),
has been proposed in Ref.~\cite{RQMD-ORG},
which is based on 
the Poincar$\acute{\rm{e}}$-invariant constrained Hamiltonian dynamics.
Although the RQMD is a numerically feasible extension of QMD
towards a fully covariant approach,
it still costs too much computing time to apply to heavy systems.
Therefore, we make an alternative extension of QMD~\cite{JAERI-RQMD1}.
We introduce Lorentz-scalar quantities into the arguments of the interactions,
\begin{eqnarray}
\label{eq:R2LS}
\bold{D}^2_{ij}
	&=& (\bold{R}_i - \bold{R}_j)^2
	\to
	\tilde{\bold{D}}^2_{ij}
	\equiv - D_{ij}^2 + (P_{ij} \cdot D_{ij})^2/P_{ij}^2\ ,\\
\label{eq:Q2LS}
\bold{Q}^2_{ij}
	&=& (\bold{P}_i - \bold{P}_j)^2
	\to
	\tilde{\bold{Q}}^2_{ij}
	\equiv - Q_{ij}^2 + (P_{ij} \cdot Q_{ij})^2/P_{ij}^2\ .
\end{eqnarray}
In these expressions, thin letters ($D_{ij}, Q_{ij}, P_{ij}$) show four vectors,
and $P_{ij}$ denotes the two-body center-of-mass four momentum
\begin{equation}
P_{ij} \equiv P_i + P_j = (E_i + E_j, \bold{P}_i + \bold{P}_j)\ .
\end{equation}
It is easy to show that the quantity $\tilde{\bold{D}}^2_{ij}$ is 
the two-body distance squared in the two-body CM system.
By these procedures we can take into account 
the primary part of the relativistic dynamical effects
approximately in our QMD.

\section{Fragment Formation}
\label{sec:NEP}

There are many attempts to describe the IMF formation
in multifragmentation processes
by using microscopic molecular dynamics (MD) simulations.
Since nuclear fragmentation is a truly quantum-mechanical process
and it is beyond the applicability limit of semiclassical MD theory,
it is necessary to improve MD theory to include quantum mechanical
effects~\cite{OR97b,OH96c,FMD,AMD-QL,OH96a,QMD-Dwidth}.
%
%
In Ref.~\cite{FMD,QMD-Dwidth},
the widths of nucleon wave packets are treated
as time-dependent dynamical variables. 
With this extension, 
we can find significant improvements in 
	ground state properties such as binding energies,
	density profiles or $\alpha$-clustering structure in light nuclei,
	nucleon emission or disintegration of clusters in medium energy 
		heavy-ion collisions.
%
%
Another type of proposed modification is to include 
quantum fluctuation or quantum branching~\cite{OR97b,OH96c,AMD-QL}. 
In Refs.~\cite{OR97b,AMD-QL}, the authors discussed the necessity of 
quantum fluctuation in the equation of motion of MD 
in order to ensure that the wave packet distribution 
follows the quantum statistics at equilibrium. 
The effects of quantum fluctuation in fragmentation processes
such as evaporation of particle and fission like fragmentation
are discussed in Ref.~\cite{AMD-QL}.
The necessity of quantum branching is discussed in Refs.~\cite{OH96c,OH96a}.
The authors introduce the quantum branching process to compensate 
the restriction on the one body nucleon distribution,
and incorporate the diffusion and the deformation of wave packets 
according to the Vlasov equation~\cite{OH96c}.
By this modification, the nucleon motion in the mean field is improved,
and the fragment mass distribution is successfully reproduced
in heavy ion reactions.

%
%
It may be possible to improve the description of 
IMF formation also in high-energy $pA$ reactions by making above modifications.
However,
all of these modifications are formulated in non-relativistic ways,
and the way to include quantum effects is under discussion.
Thus in this work, we treat the fragmentation process
in a more phenomenological manner.

In a standard treatment of fragment formation,
first we calculate the formation of excited fragments in a transport model
or in a phenomenological way,
then {\em provided that these fragments are thermalized}
they are forced to de-excite through particle emission and/or fission
in a statistical model by using only a few quantities of fragments
such as the mass, excitation energy and angular momentum.
Since we would like to address the non-equilibrium properties of
fragment formation in high-energy $pA$ reactions, 
here we do not assume that nucleons are distributed uniformly
but utilize information in the dynamical stage
--- nucleon spatial and momentum distribution inside the residual nuclei
generated in the transport model, JAM/MF.
%
In order to activate these nucleon phase space degrees of freedom
in fragmentation processes,
the percolation model would be an appropriate starting point.
In percolation models, nucleon positions are already considered,
and we can easily assign the momentum to each nucleon.

In this section, we explain our prescription to form residual nuclei
from transport model results, and how to incorporate non-equilibrium
nature in the percolation model.

\subsection{Residual nucleus formation}
\label{subsec:RNF}

After successive collisions and mean field evolution of nucleons,
several particles are emitted and a residual nucleus is formed.
Here, we take all phase space information of all particles 
consisting the system.
Then we can obtain physical property of residual nucleus such as
charge, mass and excitation energy as follows.

We assume an external nuclear mean field of nucleons 
around the residual nuclei having a depth of 50 MeV.
Then we assume that particles are trapped and form a residual nucleus
when the nucleon single particle energy is
negative (for neutrons) or below the Coulomb barrier (for protons).
%
%
The excitation energy $E^{*}$ of the residual nucleus is calculated
in a similar way to that in the exciton model~\cite{EXCITON}.
Here, we assume the nucleon Fermi energy
in the residual nuclei $\varepsilon_{F}$ is 40 MeV.     
\beqar
E^{*}
   &=&\sum_{i=1}^{N_h} \varepsilon_{i}^{h} 
    + \sum_{j=1}^{N_p} \varepsilon_{j}^{p}
\ ,\\
\varepsilon_{i}^{h}
   &=&\varepsilon_{F}-\bar{\varepsilon}_{i}
\ ,\quad
\varepsilon_{j}^{p}
    = \varepsilon_{j}-\varepsilon_{F}
\ .
\eeqar
Here $\bar{\varepsilon}_i$ denotes
the target nucleon single particle energy
just before the first collision of that target nucleon,
then the hole number ($N_h$) corresponds to the 
number of target nucleons which collide with
the leading proton or secondary cascade particles.
The particle energy $\varepsilon_j$
is the scattered nucleon single particle
energy at the time of residual nucleus formation.
The number of particles ($N_p$) is calculated
at the residual nucleus formation.
This method is basically the same as that of Ref.~\cite{SMM}.

\subsection{Non-Equilibrium Percolation model}
\label{subsec:NEP}

%
%
In (bond) percolation models,
nucleons in the residual nucleus are put on the lattice sites,
and we cut the bond between the nearest neighbor sites
according to the bond breaking probability given
as a function of excitation energy. 
Then the connected nucleons are assumed to form fragments.
By estimating the excitation energy and the mass of the residual nucleus
in a simple physical model, 
it has been demonstrated that
the percolation model works very well to describe
	the mass and energy distribution of fragments
	produced from multifragmentation processes 
	in relativistic energy ( $>$ 10 GeV ) 
	$pA$ reactions~\cite{Perc1,Perc2,Perc3,Perc4,Perc5}. 

%
%
In standard percolation models,
there is one strong assumption which disables us to describe IMF anisotropy:
the total excitation energy is assumed to be uniformly distributed
over the entire excited system of the residual nucleus,
and the nucleons are spatially distributed uniformly in a sphere.
This is equivalent to assume that equilibrium is reached in the residual nucleus.
This assumption is very convenient to discuss some formal aspects of the model,
such as the critical bond breaking probabilities, critical exponents, 
and so on.
In reality,
	the nucleon may not be uniformly distributed spatially
and	the excitation may depend on the position inside the residual nucleus,
if the fragmentation time scale is short.
Therefore, as a natural extension of the percolation model to include
non-equilibrium effects,
we start percolation
	from nucleon phase space distribution
	given in the dynamical transport model calculation.
In this case,
since nucleons are not necessarily on the lattice sites,
the bond breaking probability becomes a function of distance
between the nucleon pair.
In addition,
we can consider the position dependence of the bond breaking probability.

%
%
Here we propose a Non-Equilibrium Percolation (NEP) model
without lattice combined with a transport model.
We start from nucleon phase space distribution 
given in the microscopic transport simulation, 
and the bond breaking probability is parametrized
as a function of nucleon pair distance and positions.
%
%
In the following, we explain the actual calculation procedure in NEP.

\begin{enumerate}
\item
   Dynamical evolution of the $pA$ reaction is simulated
   by using the transport model described
   in the previous section, Sec.~\ref{sec:Transport}, 
   until $t=t_{sw}$.

\item
   Phase space information of nucleons inside the residual nucleus,
   and the bulk quantity of the residual nucleus
   such as charge $Z_{res}$, mass $A_{res}$ and excitation energy $E^{*}$ 
   are obtained according to the prescription described in the previous
   subsection, Subsec.~\ref{subsec:RNF}.   

\item
   All nucleons inside the residual nucleus are connected by bonds.

\item
   Break the bond when the bond length $L_b$ is large enough.
   When the total excitation energy exceeds a given value,
   $E^* \geq E_0 (A_{res})$ ($E_0 = 8 c_1 A_{res}$ MeV),
   the bond is broken if $L_b > R_{cut}$.
   When the total excitation energy is small, $E^* < E_0(A_{res})$, 
   bonds are broken if $L_b > c_2 R_{cut}$.
   This total excitation energy dependence is necessary
   to keep the residual nucleus stable at low excitation.

\item
   The bond between the $i$-th and $j$-th nucleons is broken
   with a probability $p_b (i,j)$. 
   The bond breaking probability $p_b (i,j)$ depends 
   on the local excitation of the residual nucleus,
   and it is parametrized as follows.
\beq
p_b (i,j)=1 - \exp\left[-\nu E^{*}(\bold{R}_{ij})\right]
\eeq
Here, $E^{*}(\bold{R}_{ij})$ is the local excitation energy
at the center-of-mass position ($\bold{R}_{ij}=(\bold{R}_i+\bold{R}_j)/2$)
of the nucleon pair.
We calculate $E^{*}(\bold{R}_{ij})$ as follows,
\beqar
E^{*}(\bold{R}_{ij})
     &=& \sum_k e^{*}_k
                \exp\left[
			-\frac{(\bold{R}_{ij} - \bold{R}_k)^2}{a^2}
			\right]
\ , \\
e^*_k
     &=& \cases{
     		t_k - e & ($t_k \geq e$)\cr
     		0       & ($t_k < e$)\cr
     	}
\ ,
\eeqar
where $t_k$ is the kinetic energy of $k$-th nucleon
in the CM frame of the residual nucleus and
$\bold{R}_k$ is the position of $k$-th nucleon.

\item
   Fragments are identified as the bond connected nucleons at this stage.
   The fragment momentum $\bold{P}_f$ and position $\bold{R}_f$
   are determined as the CM momentum and position
   of nucleons belonging to this fragment.
\end{enumerate}

The model parameters, $R_{cut},\nu,a,e,c_{1},c_{2}$, are 
determined to fit the mass distribution of fragments produced 
in p(11.5GeV)+Au reaction.
The values of NEP parameters are shown in Table~\ref{tab:NEP}. 

\begin{table}[tbh]
\caption{Parameters of NEP}
\label{tab:NEP}
\begin{center}
\begin{tabular}{c|c|c|c|c|c}
\hline 
    $R_{cut}$(fm)& $\nu$(1/MeV) & $a$(fm) & $e$(MeV) & $c_1$ & $c_2$  \\
\hline
        2.1      &    600       &  2.0    & 45.0     & 0.2   & 1.1    \\
\hline 
\end{tabular}
\end{center}
\end{table}

\begin{figure}[bht]
\begin{center}
\geteps{15cm}{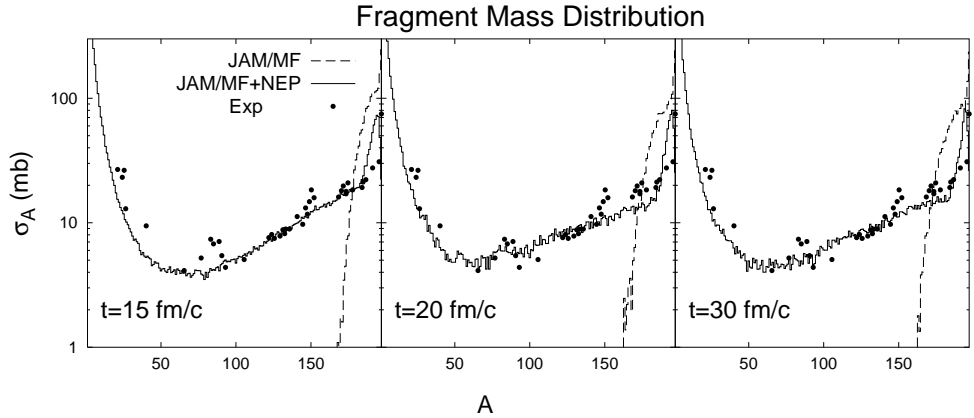}
\end{center}
\caption{
Calculated fragment mass distributions in p(12 GeV)+Au reaction.
Dashed histograms and solid histograms
show calculated results of JAM/MF and JAM/MF followed by NEP, respectively.
In the left, middle and right panels, 
the dependence on the the switching time
($t_{sw}=$15, 20, and 30 fm$/c$, respectively) is shown.
Solid points show the experimental data
of p(11.5GeV)+Au reaction~\protect{\cite{pAu11.5-300}}.
In the transport model,
simulation calculations are made
in the impact parameter range of $b \leq 7.5$ fm. 
}
\label{fig:FRGMASS}
\end{figure}

In Fig.~\ref{fig:FRGMASS},
we show the calculated fragment mass distributions in p(12 GeV)+Au reaction
in comparison with the experimental data~\cite{pAu11.5-300}.
The mass of residual nucleus (dashed histograms) is distributed
in the range $160 \lsim A_{res} \lsim 197$.
This means that, in the transport model simulation,
the number of emitted nucleons is less than 40.
%
%
Then, the decay of the residual nucleus to various mass fragments
is described in NEP as shown by the solid histograms.
One can see that the characteristic U-shape curve of
the experimental mass distribution is well reproduced
in our calculation, and this behavior does not strongly depends
on the switching time.
%
%
When we analyze the impact parameter dependence of the mass distribution,
we find that this U-shape distribution is a consequence
of the integration over the impact parameter.
At central collisions,
the residual nucleus generally has smaller mass and larger excitation energy,
and it decays into smaller fragments, in average.
At peripheral collisions,
the residual nucleus is not highly excited, and emits several nucleons 
even in the percolation stage.

\begin{figure}[bht]
\begin{center}
\geteps{15cm}{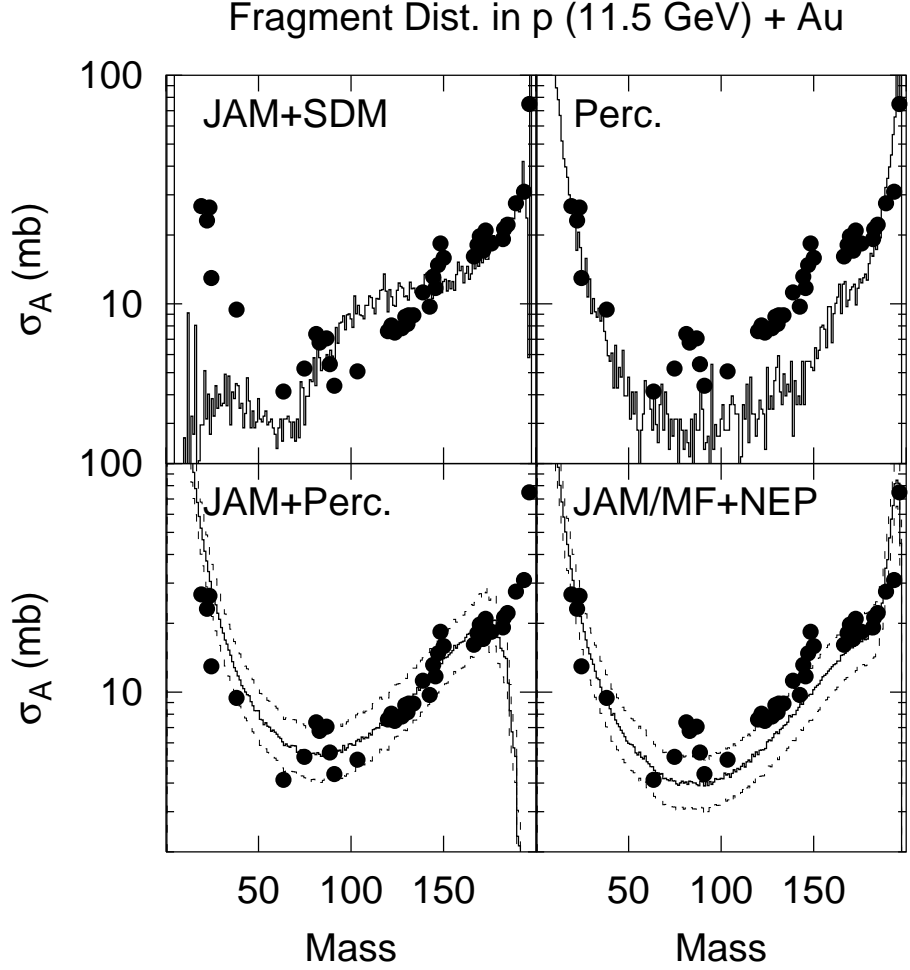}
\end{center}
\caption{
Model comparison of fragment mass distribution in p(12 GeV)+Au reaction.
Four model results ---
JAM+SDM (upper left),
Percolation model with default parameters (upper right)~\protect{\cite{Perc4}},
JAM+Percolation (bottom left), JAM/MF + NEP (bottom right)
--- are shown in histograms and compared
with the experimental data~\protect{\cite{pAu11.5-300}}.
As for the explanation of these models, see the text.
Dashed histograms in the lower panels show the calculated results
multiplied by 1.3 and 1/1.3,
showing that the combined analyses of a transport model
with percolation reproduce the data within around 30 \% accuracy.
}
\label{fig:FRGMASS2}
\end{figure}

%
%
%
It should be noted that the above global feature of the mass distribution 
can be also reproduced in a standard percolation model.
In Fig.~\ref{fig:FRGMASS2}, we show the model dependence of the
mass distribution. 
Here we compare the combination of the models,
JAM with Statistical Decay Model (SDM), 
Percolation model,
JAM with Percolation model,
and JAM/MF with NEP.
In SDM, the evaporation of light fragments
($n, p, d, t, {}^3\hbox{He}, \alpha$)
and the fission of heavy nuclei are included,
but these are not enough to form IMFs abundantly.
On the other hand, in the percolation model of Ref.~\cite{Perc4}
which fit the mass distribution of $p$+Ag reaction at 300 GeV,
the global mass distribution is reproduced without any change of 
the model parameters, and it is possible to get a better fit 
by using target dependent parameters.

This comparison of model calculations suggests the necessity
to invoke some kind of multifragmentation mechanism such as the percolation
in addition to the evaporation and fission.
If we adopt the percolation model in fragmentation stage,
we can qualitatively describe the U-shape mass distribution,
which we cannot explain in statistical decay models
with evaporation and fission.
Furthermore, in order to reproduce the whole range of the mass spectrum,
it is necessary to stabilize the nucleus at low excitation.
In the original percolation model~\cite{Perc4} (upper right panel), 
nucleon positions are kept to be on sites, and the bond breaking
probability becomes smaller in peripheral collisions,
then larger mass nuclei can survive.
With QMD transport (lower right panel),
heavy residual nuclei can survive percolation
at low excitation, because the mean field keeps the nucleus stable.
Without mean field (lower left panel), on the other hand,
nucleons with moderate kinetic energy can easily escape 
from residual nuclei, then we cannot form stable heavy nuclei.

This comparison tells us the merit to use a combined framework
of JAM/MF with percolation for 
the analysis of the fragmentation processes.
However, it is possible to reproduce the mass spectra 
without invoking spatial and/or momentum fluctuation
generated in the dynamical stage by modifying parameters.
Therefore, for the investigation of non-equilibrium nature
in the fragmentation, it is necessary to analyze 
the energy and angular distribution of fragments.

\section{Results}
\label{sec:Results}

\subsection{IMF formation points}

\begin{figure}[bth]
\begin{center}
\geteps{14cm}{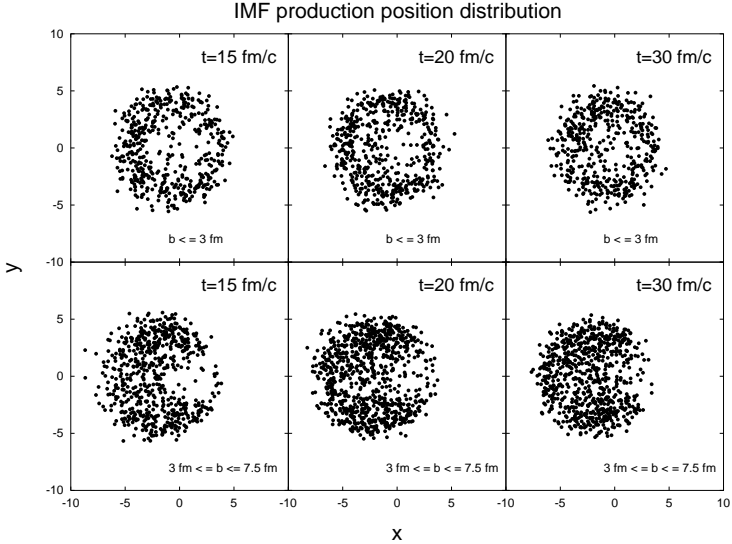}
\end{center}
\caption{
Calculated IMF (6 $\leq Z \leq$ 20) position distributions in $xy$-plain.
Solid points display the IMF formation points in NEP
following the transport model evolution 
until the switching time of 15, 20 and 30 fm$/c$.
5000 events of the transport model events are overlayed.
In upper and lower panels,
central ($b < 3$ fm) and peripheral ($3 \leq b < 7.5$ fm) events are shown,
respectively.
}
\label{fig:IMFPxy}
\end{figure}

Now we analyze the non-equilibrium properties of IMF production
in the framework explained in the previous sections.
In Fig.\ref{fig:IMFPxy}, IMF(6 $\leq Z \leq$ 20) formation position 
distributions in $xy$-plain calculated in the transport model calculation 
followed by NEP are shown.
%
In the transport model simulations, 5000 events are generated,
and the results of NEP from these events are overlayed in the figure.
In central events ($b \leq 3$ fm, upper panels), 
we can see that IMFs are produced in the doughnut shaped region.
Along the leading proton path, a large part of nucleons collide with 
the leading proton or secondary cascade particles.
Since they have large kinetic energies,
they are not judged to belong to the residual nuclei.
Even if they stay in the residual nuclei, 
the local bond breaking probability becomes large.
As a result, the IMF formation is suppressed along the leading proton path.
%
%
%
In the surrounding doughnut shaped region,
the number of collision is small, then the excitation energy is low.  
%

%
%
In the lower panels of Fig.\ref{fig:IMFPxy},
IMF formation points in peripheral collisions
(3 fm $\leq b \leq$ 7.5 fm) are shown.
We can see that IMFs are produced from a new moon shaped region.
In peripheral collisions,
highly excited region is shifted to outer side.
Then peripheral excited region shrinks to narrower at time step 30fm$/c$.    

\begin{figure}[bth]
\begin{center}
\geteps{14cm}{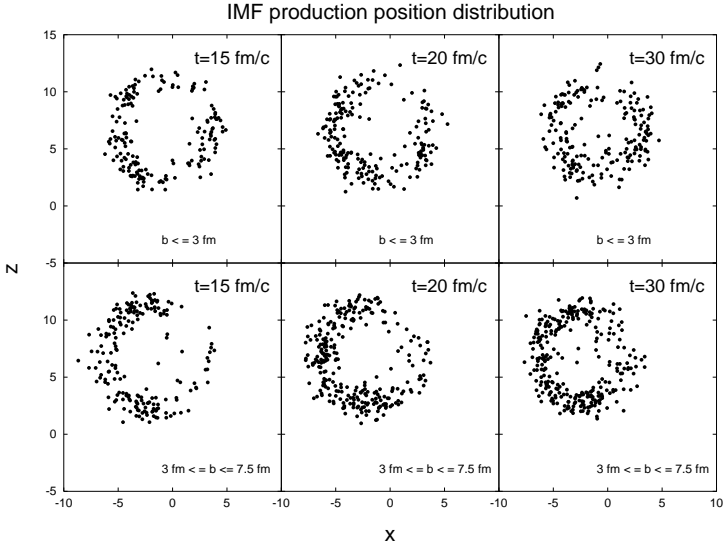}
\end{center}
\caption{
Calculated IMF (6 $\leq Z \leq$ 20) position distributions 
in $xz$-plain, restricted to $|y|$ $\leq$ 2.0 fm.
%
%
Solid points display the IMF formation points in NEP
following the transport model evolution 
until the switching time of 15, 20 and 30 fm$/c$.
5000 events of the transport model events are overlayed.
In upper and lower panels,
central ($b < 3$ fm) and peripheral ($3 \leq b < 7.5$ fm) events are shown,
respectively.
}
\label{fig:IMFPxz}
\end{figure}

In Fig.~\ref{fig:IMFPxz},
we show calculated IMF formation points in $xz$-plain,
restricted to $|y| \leq$ 2.0 fm.
We can see that IMF formation point distributions clearly 
have a hole along the leading proton path in central collisions (upper panels)
and at early times in peripheral collisions (lower panels).

In both of the cases, 
IMFs are formed mainly on the surface of the residual nuclei,
and the formation is suppressed in the central part.
This feature can be understood as follows.
When the excitation energy is small, nucleons are likely to be connected
with each other. In the central part of the residual nucleus,
IMF can grow easily to heavier nucleus by being connected with cold nucleons
in the surrounding region.
This mechanism may emerge also in the equilibrium percolation model.
However, this will not generate any anisotropy in IMF angular distribution.
%


\subsection{IMF energy distribution}

After the fragments being simultaneously produced 
through the multifragmentation of the residual nucleus,
the fragments are dispersed by the Coulomb repulsion.
In the uniform Coulomb expansion of fragments, 
the total fragment kinetic energy is calculated 
as the sum of the thermal and the Coulomb energies.
In order to describe the Coulomb expansion of fragments, 
here we consider the following classical Hamiltonian $H$,
\beqar
H=\sum_{i=1} \sqrt{\bold{P}_{fi}^2+m_{fi}^2} 
+ \sum_{i<j}\frac{Z_{f i} Z_{fj} e^2}{|\bold{R}_{fi} -\bold{R}_{fj}|} 
\label{eq:Coul}
\eeqar
where $m_{fi}$, $\bold{R}_{fi}$, $Z_{fi}$ and $\bold{P}_{fi}$ are
the mass, position, charge, and momentum of the $i$-th fragment, respectively.
From this Hamiltonian Eq.(\ref{eq:Coul}), 
the Newtonian equations of motion are solved numerically.
The initial values of the $\bold{R}_{fi}$ and $\bold{P}_{fi}$ are 
given in NEP as discussed in Sec.~\ref{sec:NEP}.
The time evolution of the Coulomb expansion is solved until
the total kinetic energy of each fragment ceases to show appreciable changes.
Then we obtain kinetic energies of fragments in the final state.
Here, we adopt the time step of $\Delta t$=0.5 fm$/c$
and integration until 500 time steps are carried out.

\begin{figure}[bth]
\begin{center}
\geteps{15cm}{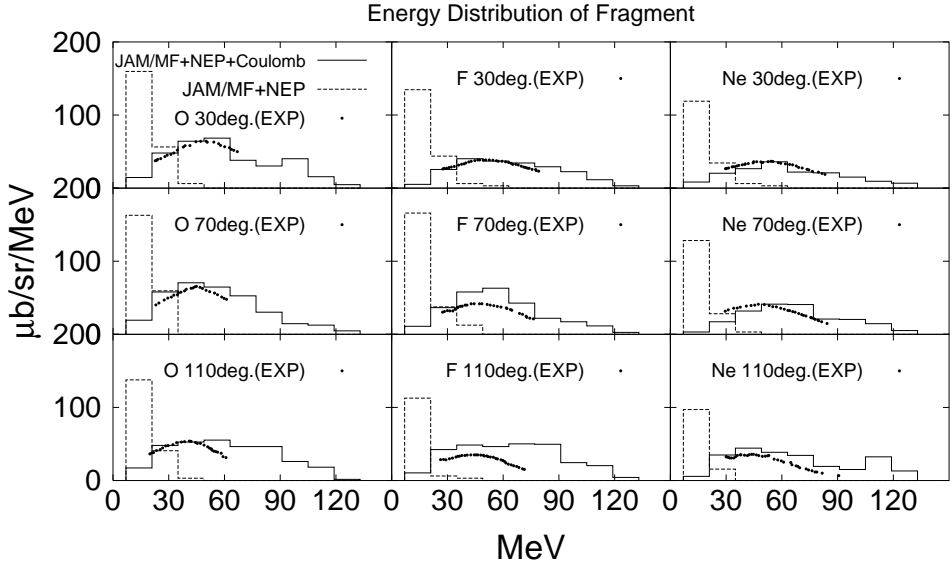}
\end{center}
\caption{
Calculated energy distributions of O, F and Ne isotopes produced in
$p$(12 GeV)+Au reaction in comparison with the experimental data.
The vertical axis shows the cross section and the horizontal axis
shows the kinetic energy of the fragment.
Left, middle and right panels show the energy distributions of O, F and Ne,
respectively.
Upper, middle and lower panels show the energy distributions emitted at 
angles of 30, 70 and 110 degrees
from incident proton direction in the laboratory frame.
Solid and dashed histograms show the calculated results with
and without the Coulomb expansion process.
The results of switching time $t_{sw} = 20$ fm$/c$ are shown.
Solid points are the experimental data~\protect\cite{ENGofpAu} 
}
\label{fig:IMFENG}
\end{figure}

In Fig.~\ref{fig:IMFENG},
we compare the calculated IMF energy distributions produced in 
$p$(12 GeV)+Au reaction with the experimental data~\cite{ENGofpAu}.
From these results we can see large effects of Coulomb expansion
in the energy distribution of IMF.
If we take into account only the kinetic energy of fragments
calculated in the transport model followed by NEP,
the energy peak around 50 MeV is never reproduced.
%
%
On the other hand, 
we find that the calculated results after Coulomb expansion 
reproduce the qualitative behavior of the experimental data well;
peaked around 40-60 MeV, enhancement at sideward angles,
and the absolute values of the double differential cross sections.
These facts show us that the Coulomb expansion process plays a very important
role in the fragmentation phenomenon.

\subsection{Analysis of IMF angular distribution}

\begin{figure}[bth]
\begin{center}
\geteps{11cm}{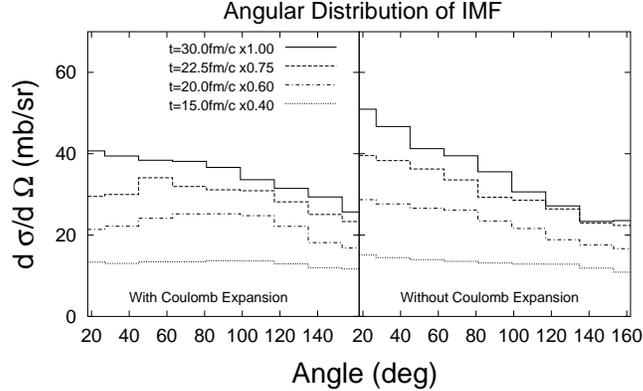}
\end{center}
\caption{
Calculated switching time dependence of the IMF angular distribution.
Dotted, dash-dotted, dashed and solid histograms 
display the results at switching time of 15, 20, 22.5 and 30 fm$/c$,
multiplied by 0.4, 0.6, 0.75 and 1.0, respectively.
In the left (right) panel, the results with (without) Coulomb expansion 
is shown.
}
\label{fig:IMFANG}
\end{figure}

In Fig.~\ref{fig:IMFANG},
we show the calculated switching time dependence
of the IMF angular distribution.
%
%
%
In comparing the results with (left panel) and without (right panel)
Coulomb expansion,
the Coulomb expansion effects are found to modify the angular distribution
largely from forward peaked to sideward peaked.
This mechanism is connected with the IMF formation point 
in the residual nucleus.
As shown at Fig.~\ref{fig:IMFPxy} and Fig.~\ref{fig:IMFPxz},
IMFs are produced dominantly 
in a doughnut shaped region in violent central collisions.
Then the Coulomb force among fragments pushes IMFs sideways
of the doughnut region
even if IMFs are forward peaked before the Coulomb expansion.
In peripheral collisions,
although IMFs are formed in a new moon shaped regions at early times,
the rapid energy transport washes out the anisotropy
of the IMF production points. 
%
%
Then if the impact parameters are integrated over, 
the angular distribution of IMFs becomes forward peaked 
at around the switching time $t_{sw}=30$ fm$/c$
even after the Coulomb expansion as shown
in the left panel of Fig.~\ref{fig:IMFANG}.
This observation is consistent with 
the behavior expected in a equilibrium statistical picture of fragmentation.
Therefore,
our present study suggests the possibility
that IMF angular distribution would have sideward peak
at around 90 degrees in the rest frame of the residual nuclei
if IMFs can be formed in a very short time scale around 20 fm$/c$.

\section{Summary}
\label{sec:Summary}

In this paper, 
we have analyzed the mechanism of formation and sideward enhancement of IMFs
produced in the 12 GeV proton induced reaction on Au target
by using a combined framework of a transport model (JAM/MF) 
followed by a newly developed Non-Equilibrated Percolation (NEP) model.
NEP can take into account
	non-uniform spatial distribution of nucleons
and	the local excitation in the residual nucleus,
then it would be appropriate to analyze non-equilibrium natures of 
IMF formation.
The present combined model has been verified in various ways.
In the hadronic cascade stage (JAM), 
	momentum distribution of nucleons has been verified
	by comparing hadron (proton,$\pi^{+}$,$\pi^{-}$,$K^{+}$,$K^{-}$)
        transverse mass spectra~\cite{JAM}.
The effective interaction in QMD has been verified
	through the study of matter and nuclear properties
	and the quasi-elastic neutron energy spectra in $pA$ reactions at
	256 MeV and 800 MeV~\cite{QMD-MEANB}.
In the fragmentation (NEP),
	although there are still several model parameters, 
	they are determined to fit the fragment mass spectra
	in high-energy $pA$ reactions.
By using the same parameter, 
	we applied our model to the fragment energy distributions,
	and the experimental data are well reproduced.

%
%
By using this model we have shown that IMFs are produced 
	in a doughnut shaped region and a new moon shaped one
	inside the residual nucleus in central and peripheral collisions,
	respectively.
We suggest that combined effects of such a geometry of 
	IMF formation point distribution
and	the Coulomb expansion between fragments
would lead to a sideward peaked angular distribution of IMFs
provided that the IMFs are formed in a very short time scale
of around 20 fm$/c$.
In principle, it would be possible to estimate this time scale experimentally
by measuring the particle correlations in momentum multi-dimensions.
%
%
Another remaining interesting problem is to study 
	the mechanism of the incident energy dependence
	of the IMF angular distribution,
	which evolves from forward peaked, sideward peaked, to 
	backward peaked as the incident energy grows.
Our approach will be also useful in analyzing those problems.

\section*{Acknowledgements}
\label{sec:Ack}

The authors are grateful to 
Prof. K.H. Tanaka and Dr. T. Murakami for preparing the experimental data
and valuable comments and discussions.
We would like to thank Prof. W. Bauer for providing us
of the computer program of the original percolation model.
It is a pleasure to acknowledge the hospitality and useful discussions 
of the members of the research group 
for nuclear theory at Hokkaido university
and Hadron Science at Japan Atomic Energy Research Institute.
The authors also would like to acknowledge Prof. Tomoyuki Maruyama
for useful comments and discussions.


\appendix

\section{Quantum molecular dynamics ( QMD )}
\label{app:QMD}

In QMD method, the single particle wave function of the $i$-th nucleon
is represented in a Gaussian wave packet, having the density distribution, 
\beqar
\rho_{i} (\bold{r}) = \frac{1}{(2 \pi L)^{\frac{3}{2}}}
                      e^{-\frac{(\bold{r}-\bold{R}_i)^{2}}{2L}},
\eeqar
where $L$ denotes the width parameter for the Gaussian wave packets and
$\bold{R}_i$ are the centers of position of the $i$-th nucleon.

Each term in the interaction energy in Eq.~(\ref{eq:effect}) depends
on the center of position $\bold{R}$ of the nucleon and 
the center of momentum $\bold{P}$ of the nucleon as follows,
\beqar
V_{\rm{Pauli}}
   &=& \frac{C_{Pauli}}{2 (q_0 p_0 / \hbar )^3 }\sum_{i \ne j}
	e^{
	  -\frac{(\bold{R}_i-\bold{R}_{j})^2}{2q^2_{0}}
	  -\frac{(\bold{P}_i-\bold{P}_{j})^2}{2p^2_{0}}
	  }
	\delta_{\tau_i \tau_j} \delta_{\sigma_i \sigma_j}, \\
V_{\rm{Skyrme}}
   &=&	 \frac{\alpha}{2\rho_{0}}\sum_i <\rho_i>
	+\frac{\beta}{(1+\tau)\rho^{\tau}_0} \sum_i <\rho_i>'^{\tau}, \\
V_{\rm{Sym}}
   &=&	\frac{V_{sym}}{2\rho_0}\sum_{i \ne j}
	(1-2|c_i-c_j|)\rho_{ij}, \\
V_{\rm{Mom}}
   &=&	\frac{1}{2\rho_{0}} 
	\sum_{i \ne j}
	\left[
	  	\frac{V^{(1)}_{ex}}
		{1+\left[ \frac{\bold{P}_i-\bold{P}_j}{\mu_1 \hbar} \right]^2} 
  		+
		\frac{V^{(2)}_{ex}}
		{1+\left[ \frac{\bold{P}_i-\bold{P}_j}{\mu_2 \hbar} \right]^2}
	\right]
	\rho_{i j}, \\
V_{\rm{Coulomb}}
   &=& \sum_{i \ne j}
	\frac{c_i c_j e^2}{2}
	\frac{1}{|\bold{R}_i -\bold{R}_j|}
	\rm{erf}(\frac{|\bold{R}_i-\bold{R}_j|}{\sqrt{4L}}).      
\eeqar
Here, erf denotes the error function, 
and $c_{i}$ is 1 for protons and 0 for neutrons. 
We only take into account the effective interactions which affect nucleons.
$<\rho_{i}>$ is an overlap of density with other nucleons defined as 
\beqar
<\rho_{i}> 
           \nonumber
           &\equiv& \sum_{j ( \ne i )} \rho_{i j}
           \equiv \sum_{j (\ne i)}
	   \int d \bold{r} \rho_i (\bold{r}) \rho_j (\bold{r})\\
           &=& \sum_{j (\ne i)}  
           \frac{1}{(4 \pi L)^{\frac{3}{2}}}
	   e^{-\frac{(\bold{R}_i - \bold{R}_j)^2}{4L}}.
\eeqar
$<\rho_{i}>'$ means that $L$ of $<\rho_{i}>$ is modified to be 
$L'=\frac{1}{2}(1+\tau)^{\frac{1}{\tau}} L$.                    
The parameters are listed in Table~\ref{tab:QMDpara}.

\begin{table}[h]
\caption{QMD parameters}
\label{tab:QMDpara}
\begin{center}
\begin{tabular}{|cc|cc|}
\hline
  $C_{\rm{Pauli}}$  &    140     $[\rm{MeV}]$     
& $V_{\rm{sym}}$    &    25.0    $[\rm{MeV}]$
\\\hline
  $q_0$             &      1.644 $[\rm{fm}]$      
& $V^{(1)}_{ex}$    &  -258.54   $[\rm{MeV}]$
\\\hline
  $p_0$             &    120     $[\rm{MeV/c}]$   
& $\mu_{1}$         &     2.35   $[\rm{fm^{-1}}]$
\\\hline
  $L$               &      1.75  $[\rm{fm^2}]$
& $V^{(2)}_{ex}$    &   375.6    $[\rm{MeV}]$
\\\hline
  $\alpha$          &   -127.86  $[\rm{MeV}]$
& $\mu_{2}$         &     0.4    $[\rm{fm^{-1}}]$
\\\hline
  $\beta$           &    204.28  $[\rm{MeV}]$
& $\rho_{0}$        &     0.168  $[\rm{fm^{-3}}]$
\\\hline
  $\tau$            &    $\frac{4}{3}$
&                   &                             \\    
\hline
\end{tabular}
\end{center}
\end{table}

\end{document}